\documentclass[%
 preprint,
 amsmath,amssymb,
 aps,
]{revtex4-2}
\usepackage{amsmath}
\usepackage{graphicx}
\usepackage{dcolumn}
\usepackage{bm}
\usepackage{hyperref}

\begin{document}


\title{A model-independent irreducible tensor formalism to analyse the sequential decays of the triple strange process $\bar{p} p \rightarrow \bar{\Omega}^{+} \Omega^{-} \rightarrow K^{+} \bar{\Lambda} K^{-} \Lambda \rightarrow K^{+} \bar{p} \pi^{+} K^{-} p \pi^{-}$ }

\author{Deepak Pachattu}
\affiliation{Department of Physics, BITS Pilani K K Birla Goa Campus, Zuarinagar 403726, Goa, India}
\begin{abstract}
A model-independent irreducible tensor formalism is developed to analyse $\bar{\Omega}\Omega$ production in $\bar{p}p$ collisions and its subsequent decay via the triple-strange decay process, $\bar{\Omega}^{+} \Omega^{-} \rightarrow K^{+} \bar{\Lambda} K^{-} \Lambda \rightarrow K^{+} \bar{p} \pi^{+} K^{-} p \pi^{-}$. These processes are relevant for the upcoming experiments at PANDA. We not only provide expressions for the density matrix of the $\Omega\bar{\Omega}$ system but also for the products at each stage of the decay. The Fano statistical tensors so obtained not only completely characterize the relevant final systems but also provide expressions for joint angular distributions.  Finally, we show which of the Fano statistical tensors characterizing the $\bar{\Omega}\Omega$ system can be inferred from the Fano statistical tensors characterizing the final $\bar{p}p$ system, wherein we also obtain the relation between the production cross sections for $\Omega\bar{\Omega}$ and the final $p\bar{p}$ system. 
\end{abstract}
\maketitle
\section{Introduction}
In view of the recent detailed simulation study of the signatures from the sequential decays of the triple-strange process: $\bar{p} p \rightarrow \bar{\Omega}^{+} \Omega^{-} \rightarrow K^{+} \bar{\Lambda} K^{-} \Lambda \rightarrow K^{+} \bar{p} \pi^{+} K^{-} p \pi^{-}$ by PANDA \cite{PANDA}, we develop a model-independent irreducible tensor formalism to analyse all the observables which are relevant for such a kinematically complete experiment. We not only provide expressions for the density matrix for the $\Omega\bar{\Omega}$ system, but also for density matrices relevant at each stage of the decay. We also indicate how some of the Fano statistical tensors which characterize the $\Omega\bar{\Omega}$ state can be determined by measuring the Fano statistical tensors of the final $p\bar{p}$ system and hence show how the production cross sections for $\Omega\bar{\Omega}$ and the final $p\bar{p}$ system are related.

A study of the density matrices for $\Omega$ production and its subsequent decays, in the context of PANDA was carried out by Thom\'{e} \cite{thome}. Thom\'{e}'s work focussed mainly on one of the decay chains, which is, that of $\Omega$ and the spin structure for $\Omega\bar{\Omega}$ production was not discussed. A polarization analysis for hyperon production and their subsequent decays in the context of $e^+e^-$ annihilation was done by Perotti et. al, \cite{perotti} which was later used by Ablikim et. al., \cite{ablikim} for a model-independent determination of the spin of the $\Omega^-$. The joint angular distribution of the decay chain $J / \psi \rightarrow\left(\Lambda \rightarrow p \pi^{-}\right)\left(\bar{\Lambda} \rightarrow \bar{p} \pi^{+}\right)$, again in the context of $e^+e^-$ annihilation was derived by F\"aldt and Kupsc \cite{faldt}, which was later used my Ablikim et. al., \cite{ablikim2} to study polarization and entanglement in baryon-antibaryon Pair Production in $e^+e^-$ annihilation. But none of these works analyse triple strange processes in $p\bar{p}$ collisions, which are relevant for PANDA. And, as the authors in \cite{PANDA} put it, ``The $\Omega^-$ hyperon on the other hand has never been studied with antiproton beams before and the production cross section is unknown.''. Further, we have not come across any work which explicitly writes down the complete spin structure for the reaction, $\bar{p} p \rightarrow \bar{\Omega}^{+} \Omega^{-}$. It is, therefore, in this context, that our formalism becomes important and relevant for PANDA. 

\section{Formalism}
\subsection{$\bar{p} p \rightarrow \bar{\Omega}^{+} \Omega^{-}$} 
The spin structure for the reaction of the type $a+A\to b+B$ involving particles/reactants $a,A$ and products $b,B$ with respective spins $s_a,s_A$ and $s_b,s_B$  can be studied by defining a matrix $\mathbf{M}$ of dimension $\left(2 s_b+1\right)\left(2 s_B+1\right) \times\left(2 s_a+1\right)\left(2 s_A+1\right)$ in the spin-space of the participants. The matrix elements are related to the on-energy-shell transition matrix $\mathbf{T}$ through
$$
\left\langle s_b \mu_b s_B \mu_B|\mathbf{M}| s_a \mu_a s_A \mu_A\right\rangle=\sqrt{\frac{2 \pi D}{v}}\left\langle s_b \mu_b s_B \mu_B ; \overrightarrow{p_f}|\mathbf{T}| \overrightarrow{p_i} ; s_a \mu_a s_A \mu_A\right\rangle
$$
where $\overrightarrow{p_f}$ and $\overrightarrow{p_i}$ are the final and initial c.m. momenta respectively. The density of final states and the magnitude of the relative velocity in the initial state are denoted by $D$ and $v$ respectively. Introducing channel spins $s_i$ and $s_f$ and partial wave expansions in the entrance and exit channels, we have,
\begin{equation}
\begin{aligned}
\label{partial}
\left\langle s_f \mu_f ; \vec{p}_f|\mathbf{T}| \vec{p}_i ; s_i \mu_i\right\rangle= & \sum_{l_f, m_f} \sum_{l_i, m_i} \sum_{j, j^{\prime}} C\left(l_f s_f j^{\prime} ; m_f \mu_f m^{\prime}\right) C\left(l_i s_i j ; m_i \mu_i m\right) \\
&\times Y_{l_{f}m_f}\left(\hat{p}_f\right)\left\langle\left(l_f s_f\right) j^{\prime} m^{\prime}|\mathbf{T}|\left(l_i s_i\right) j m\right\rangle Y_{l_i m_i}^*\left(\hat{p}_i\right)
\end{aligned}
\end{equation}
where $C(j_1,j_2j;m_1m_2m)$ denote the Clebsch-Gordan coefficients in the notation of Rose \cite{rose} and the rotational invariance implies
$$
\left\langle\left(l_f s_f\right) j^{\prime} m^{\prime}|\mathbf{T}|\left(l_i s_i\right) j m\right\rangle=\delta_{j j'} \delta_{m m^{\prime}} T_{l_f s_f; l_i s_i}^j
$$
for the energy-dependent partial-wave amplitudes, $T_{l_f s_f; l_i s_i}^j$. Using the short hand notation
$$
\left(A^k \otimes B^{k^{\prime}}\right)_Q^K=\sum_{q=-k}^k C\left(k k^{\prime} K, q q^{\prime} Q\right) A_q^k B_{q'}^{k^{\prime}}
$$
to denote an irreducible tensor of rank $K$ constructed out of two irreducible tensors $A_q^k$ and $B_{q'}^{k'}$ of ranks $k$ and $k^{\prime}$ respectively and using Racah techniques, eq. (\ref{partial}) may be rewritten as \cite{fesh}
\begin{equation}
\begin{aligned}
\label{T}
\left\langle s_f \mu_f ; \vec{p}_f|\mathbf{T}| \vec{p}_i ; s_i \mu_i\right\rangle= & \sum_{l_f, l_i, j, \lambda}(-1)^{l_i+s_i+l_f-j} W\left(s_i l_i s_f l_f ; j \lambda\right)[j]^2[\lambda]\left[s_f\right]^{-1} T_{l_f s_f;l_i s_i}^j \\
&\times C\left(s_i \lambda s_f ; \mu_i \mu \mu_f\right)(-1)^\mu\left(Y_{l_f}\left(\hat{p}_f\right) \otimes Y_{l_i}\left(\hat{p}_i\right)\right)_{-\mu}^\lambda,
\end{aligned}
\end{equation}
where we use the shorthand notation
$$
[j]\equiv \sqrt{(2j+1)}.
$$
We now express $\mathbf{M}$ as
\begin{equation}
\label{coll}
\mathbf{M}=\sum_{s_f, \mu_f} \sum_{s_i, \mu_i} \sqrt{\frac{2 \pi D}{v}}\left|s_f \mu_f\right\rangle\left\langle s_f \mu_f ; \overrightarrow{p_f}|\mathbf{T}| \overrightarrow{p_i} ; s_i \mu_i\right\rangle\left\langle s_i \mu_i\right|
\end{equation}
and use eq. (\ref{T}) for the matrix elements. Since $\left|s_f \mu_f\right\rangle$ transforms under rotations as an irreducible tensor, $K_{\mu_f}^{s_f}$, of rank $s_f$; while $\left\langle s_i \mu_i\right|$ does so like $(i)^{2 \mu_i} B_{-\mu_i}^{s_i}$, an irreducible tensor of rank $s_i$ \cite{sch}, so that
\begin{equation}
\label{proj}
\left|s_f \mu_f\right\rangle\left\langle s_i \mu_i\right|=\sum_{\lambda=|s_i-s_f|}^{(s_i+s_f)} C\left(s_f s_i \lambda ; \mu_f-\mu_i \mu\right)(-1)^{\mu_i}\left(K^{s_f} \otimes B^{s_i}\right)_\mu^\lambda.
\end{equation}
This allows us to define irreducible spin transition tensors $\mathcal{S}_\mu^\lambda$ connecting the spin spaces of $s_i$ and $s_f$ through
\begin{equation}
\label{proj-channel}
\mathcal{S}_\mu^\lambda\left(s_f, s_i\right)=(-1)^{s_i}\left[s_f\right]\left(K^{s_f} \otimes B^{s_i}\right)_\mu^\lambda,
\end{equation}
where $\lambda$ ranges from $\left|s_f-s_i\right|$ to $\left(s_f+s_i\right)$. The collision matrix, eq. (\ref{coll}), can be written on using eqs. (\ref{proj}) and (\ref{proj-channel}), in the simple {\em invariant} form \cite{grmsv}
\begin{equation}
\label{invariant}
\mathbf{M}=\sum_{s_f, s_i, \lambda}\left(\mathcal{S}^\lambda\left(s_f, s_i\right) \cdot T^\lambda\left(s_f, s_i\right)\right),
\end{equation}
where
$$
\begin{aligned}
T_{-\mu}^\lambda\left(s_f, s_i\right)= \sum_{l_f, l_i, j}(-1)^{l_i+s_i+l_f-j} W\left(s_i l_i s_f l_f ; j \lambda\right) M_{l_f s_f ; l_i s_i}^j[j]^2\left[s_f\right]^{-1} 
\left(Y_{l_f}\left(\hat{p}_f\right) \otimes Y_{l_i}\left(\hat{p}_i\right)\right)_{-\mu}^\lambda
\end{aligned}
$$
are the irreducible tensor reaction amplitudes in the channel-spin space and
$$
M_{l_f s_f ; l_i s_i}^j=\sqrt{\frac{2 \pi D}{v}} T_{l_f s_f; l_i s_i}^j.
$$
Using \cite{grmsv}
\begin{equation}
\begin{aligned}
\mathcal{S}_\mu^\lambda\left(s_f, s_i\right)=\sum_{\lambda_1, \lambda_2}\left\{\begin{array}{l}
s_b s_B s_f \\
s_a s_A s_i \\
\lambda_1 \lambda_2 \lambda
\end{array}\right\} \frac{\left[s_f\right]^2\left[s_i\right]\left[\lambda_1\right]\left[\lambda_2\right]}{\left[s_b\right]\left[s_B\right]}(-1)^{s_i-s_a-s_A}
\left(\mathcal{S}^{\lambda_1}\left(s_b, s_a\right) \otimes \mathcal{S}^{\lambda_2}\left(s_B, s_A\right)\right)_\mu^\lambda
\end{aligned}
\end{equation}
in eq. (\ref{invariant}), we obtain
\begin{equation}
\mathbf{M}=\sum_{\lambda_1,\lambda_2,\lambda}\left(\left(\mathcal{S}^{\lambda_1}\left(s_b, s_a\right) \otimes \mathcal{S}^{\lambda_2}\left(s_B, s_A\right)\right)^\lambda\cdot M^\lambda(\lambda_1,\lambda_2)\right),
\end{equation}
where the new set of irreducible tensor reaction amplitudes are related to the channel-spin amplitudes via
\begin{equation}
M^\lambda_\mu(\lambda_1,\lambda_2)=\sum_{s_f,s_i}\left\{\begin{array}{l}
s_b s_B s_f \\
s_a s_A s_i \\
\lambda_1 \lambda_2 \lambda
\end{array}\right\} \frac{\left[s_f\right]^2\left[s_i\right]\left[\lambda_1\right]\left[\lambda_2\right]}{\left[s_b\right]\left[s_B\right]}(-1)^{s_i-s_a-s_A}\, T^\lambda_\mu(s_f,s_i)
\end{equation}
Specializing the above formalism for the case of $\overline{\mathrm{p}} \mathrm{p} \rightarrow \bar{\Omega} \Omega$ involves choosing the spins to be $s_a=s_A=1/2$ and $s_b=s_B=3/2$, with $s_i=0,1$ and $s_f=0,1,2,3$. We hence obtain for
$\bar{p} p \rightarrow \bar{\Omega}^{+} \Omega^{-}$:
\begin{equation}
\label{M}
\mathbf{M}=\sum_{\lambda_1, \lambda_2=1}^2\sum_{\lambda=|\lambda_1-\lambda_2|}^{(\lambda_1+\lambda_2)}\left(\left(S^{\lambda_1}\left({\textstyle{\frac{3}{2}}},{\textstyle{\frac{1}{2}}}\right) \otimes S^{\lambda_2}\left({\textstyle{\frac{3}{2}}},{\textstyle{\frac{1}{2}}}\right)\right)^\lambda \cdot M^\lambda\left(\lambda_1, \lambda_2\right)\right),
\end{equation}
where the irreducible reaction amplitudes $M^\lambda_\mu(\lambda_1,\lambda_2)$ now have the following partial-wave expansion 
\begin{equation}
\begin{aligned}
M_\mu^\lambda\left(\lambda_1, \lambda_2\right)&= \sum_{l_i, l_f}\sum_{s_i=0}^1\sum_{s_f=0}^3\sum_j(-1)^{l_i+l_f-j+1} M_{l_f s_f ;l_i s_i}^j(E)\left(Y_{l_f}\left(\hat{p}_f\right) \otimes Y_{l_i}\left(\hat{p}_i\right)\right)_\mu^\lambda
 W\left(s_i l_i s_f l_f ; j \lambda\right)\\&\times[j]^2\frac{\left[s_f\right]\left[s_i\right]\left[\lambda_1\right]\left[\lambda_2\right]}{4}\left\{\begin{array}{ccc}
{\textstyle{\frac{3}{2}}} & {\textstyle{\frac{3}{2}}} & s_f \\
{\textstyle{\frac{1}{2}}} & {\textstyle{\frac{1}{2}}} & s_i \\
\lambda_1 & \lambda_2 & \lambda
\end{array}\right\}.
\end{aligned}
\end{equation}

The spin-state of the initial $\bar{p} p$ system is characterized by the density matrix
\begin{equation}
\label{rhoi}
\rho^i=\frac{1}{4} \sum_{k_1, k_2=0}^1 \sum_{k=\left|k_1-k_2\right|}^{\left(k_1+k_2\right)}(-1)^{k_1+k_2-k}\left(\left(P^{k_1} \otimes Q^{k_2}\right)^k \cdot\left(\sigma^{k_1}(1) \otimes \sigma^{k_2}(2)\right)^k\right),
\end{equation}
in terms of the irreducible components 
\begin{equation}
\begin{aligned}
P_0^0 & =Q_0^0=1 ; P_0^1=P_z ; Q_0^1=Q_z \\
P_{ \pm 1}^1 & =\mp \frac{1}{\sqrt{2}}\left(P_x \pm \mathrm{i} P_y\right) ; Q_{ \pm 1}^1=\mp \frac{1}{\sqrt{2}}\left(Q_x \pm \mathrm{i} Q_y\right)
\end{aligned}
\end{equation}
of the beam and target polarization vectors, $\overrightarrow{P}$ and $\overrightarrow{Q}$ and the Pauli spin-operators
$\sigma^0_0(i)=1$ and $\sigma^1_{\pm 1}(i)=\mp \frac{1}{\sqrt{2}}\left(\sigma_x(i) \pm \mathrm{i} \sigma_y(i)\right)$, with $i=1,2$, for the proton and the antiproton.
The density matrix for the final state, $\rho_{\Omega,\bar{\Omega}}$ is given by
\begin{equation}
    \rho_{\Omega,\bar{\Omega}}=\mathbf{M}\rho^i\mathbf{M}^\dagger.
\end{equation}
Using eqs. (\ref{M}) and (\ref{rhoi}) and after using the recoupling equations in \cite{grmsv}, we get
\begin{equation}
\begin{aligned}
\label{rhof}
    \rho_{\Omega,\bar{\Omega}}=\sum_{k_1'',k_2''=0}^3\sum_{k''=|k_1''-k_2''|}^{(k_1''+k_2'')}\left( t^{k''}(k_1'',k_2'')\cdot\left(S^{k_1''}\left({\textstyle{\frac{3}{2}}},{\textstyle{\frac{3}{2}}}\right)\otimes S^{k_2''}\left({\textstyle{\frac{3}{2}}},{\textstyle{\frac{3}{2}}}\right)\right)^{k''}\right)
\end{aligned}
\end{equation}
where the irreducible Fano statistical tensors $t^{k''}_{q''}(k_1'',k_2'')$ which {\em completely} describe the polarization state of the $\Omega\bar{\Omega}$ system are given by, after recoupling, as
\begin{equation}
\begin{aligned}
\label{Fano-omega}
    & t^{k''}_{q''}(k_1'',k_2'')=8 \sum_{k_1,k_2,k}\sum_{\lambda_1,\lambda_2,\lambda}\sum_{\lambda_1',\lambda_2',\lambda'}\sum_{k_1',k_2'=1}^2\sum_{k'=|k_1'-k_2'|}^{(k_1'+k_2')}
    \left(\left (P^{k_1}\otimes Q^{k_2}\right)^k\otimes\left(M^\lambda(\lambda_1,\lambda_2)\otimes M^{\dagger^{\lambda'}}(\lambda_1',\lambda_2')\right)^{\lambda''}\right)^{k''}_{q''}\\
    &\times\left\{\begin{array}{lll}
\lambda_1 & \lambda_2 & \lambda \\
k_1 & k_2 & k \\
k_1^{\prime} & k_2^{\prime} & k^{\prime}
\end{array}\right\}
\left\{\begin{array}{lll}
k_1' & k_2' & k' \\
\lambda_1^{\prime} & \lambda_2^{\prime} & \lambda^{\prime}\\
k_1'' & k_2'' & k'' 
\end{array}\right\}
W({\textstyle{\frac{1}{2}}}k_1{\textstyle{\frac{3}{2}}}\lambda_1;{\textstyle{\frac{1}{2}}}k_1')
\,W({\textstyle{\frac{1}{2}}}k_2{\textstyle{\frac{3}{2}}}\lambda_2;{\textstyle{\frac{1}{2}}}k_2')
\,W({\textstyle{\frac{3}{2}}}\lambda_1'{\textstyle{\frac{3}{2}}}k_1';{\textstyle{\frac{1}{2}}}k_1'')\\
&\times W({\textstyle{\frac{3}{2}}}\lambda_2'{\textstyle{\frac{3}{2}}}k_2';{\textstyle{\frac{1}{2}}}k_2'')
\,W(k\lambda k''\lambda';k'\lambda'')(-1)^{\lambda_1+\lambda_2}(-1)^{k_1''+k_2''}(-1)^{k+k'-k''}
[\lambda_1][\lambda_2][\lambda_1'][\lambda_2'][k_1''][k_2'']\\
&\times [\lambda][\lambda'][\lambda''][k_1][k_2][k]
[k_1']^2[k_2']^2[k']^2.
\end{aligned}
\end{equation}
From eqs. (\ref{rhof}) and (\ref{Fano-omega}), we see that the highest rank of the Fano statistical  tensor is 6, but for it to be realized, we should have both the beam and the target polarized. If we have an unpolarized beam and target, as is the case for PANDA, the maximum rank of the Fano tensor would be 4 and the Fano statistical tensors become
\begin{equation}
\begin{aligned}
t^{k}_{q}(k_1,k_2)&=\sum_{n_1,n_2,n}\sum_{n_1',n_2',n'}   \left(M^n(n_1,n_2)\otimes M^{\dagger^{n'}}(n_1',n_2')\right)^{k}_{q}
(-1)^{n_1+n_2-n}(-1)^{k_1+k_2-k}[n_1][n_2][n]\\&\times [n_1'][n_2'][n'][k_1][k_2]
W({\textstyle{\frac{3}{2}}}n_1'{\textstyle{\frac{3}{2}}}n_1;{\textstyle{\frac{1}{2}}}k_1)W({\textstyle{\frac{3}{2}}}n_2'{\textstyle{\frac{3}{2}}}n_2;{\textstyle{\frac{1}{2}}}k_2)
\left\{\begin{array}{lll}
n_1 & n_2 & n \\
n_1' & n_2' & n'\\
k_1 & k_2 & k
\end{array}\right\},
\end{aligned}
\end{equation}
where the relabeling of the free and dummy indices has been done for later convenience.

To make use of the parity conservation in this reaction, we choose a frame in which the quantization axis is parallel to $\overrightarrow{p_i} \times \overrightarrow{p_f}$ and the $x$-axis  parallel to $\overrightarrow{p_i}$ (referred to as TF, the Transverse Frame), using which we would have \cite{grmsv}
$$
Y_{l_{f}m_{f}}\left(\hat{p}_f\right)=Y_{l_{f}m_{f}}\left(\frac{\pi}{2}, \theta\right) ; Y_{l_i m_i}\left(\hat{p}_i\right)=Y_{l_i m_i}\left(\frac{\pi}{2}, 0\right).
$$
Using parity conservation, which implies $(-1)^{l_i}=(-1)^{l_f}$ and the relation
$$
Y_{l m}(\theta, \phi)=\sqrt{\frac{(2 l+1)(l-m) !}{4 \pi(l+m) !}} P_l^m(\cos \theta) e^{i m \phi}, m \geq 0,
$$
where $P_l^m$, the associated Legendre polynomials obey $P_l^m(0)=$ 0 for odd $(l-m)$, we get
\begin{equation}
\label{parity}
M_\mu^\lambda\left(\lambda_1, \lambda_2\right)=0 {\text{ for all odd }} \mu. 
\end{equation}
Using eq. (\ref{parity}) in eq. (\ref{Fano-omega}) for the case of unpolarized beam and target (i. e., $k_1=k_2=k=0$ in eq. (\ref{Fano-omega})), we get the condition
\begin{equation}
\label{Fano-symm}
t^{k}_{q}(k_1,k_2)=0 \text{ for all odd } q,
\end{equation}
for the Fano statistical tensors.

The angular distribution (diffrential cross section) of the $\Omega$($\bar{\Omega})$ can be found by taking the trace of eq. (\ref{rhof}). If the beam and target are unpolarized, we get
\begin{equation}
\label{unpol-dcs}   \mathrm{Tr}\left(\rho_{\Omega,\bar{\Omega}}\right)=16\sum_{\lambda_1,\lambda_2,\lambda}\sum_{\mu=-\lambda}^\lambda \left|M^\lambda_\mu(\lambda_1,\lambda_2\right|^2.
\end{equation}
where we have used the normalization \cite{grumer},
\begin{equation}
\left\langle s m^{\prime}\left|S_q^k(s,s)\right| s m\right\rangle=C\left(s k s ; m q m^{\prime}\right)[k].
\end{equation}
Using eq. (\ref{parity}), eq. (\ref{unpol-dcs}) reduces to
\begin{equation}
\label{unpol-dcs2}   \mathrm{Tr}\left(\rho_{\Omega,\bar{\Omega}}\right)=16\sum_{\lambda_1,\lambda_2,\lambda}
\sum_{\text{even }\mu} \left|M^\lambda_\mu(\lambda_1,\lambda_2\right|^2,
\end{equation}
where the possible even values of $\mu$ are:\\
$\mu=0$, when $\lambda=0,1,2,3,4$\\
$\mu=2$, when $\lambda=2,3,4$ and\\
$\mu=4$, when $\lambda=4$.
\subsection{The first stage of decay: the density matrix for $\Lambda\bar{\Lambda}$}
The density matrix $\rho_{\Lambda\bar{\Lambda}}$ is given by
\begin{equation}
\begin{aligned}
\label{rhoLambda}
    \rho_{\Lambda,\bar{\Lambda}}=\sum_{k_1,k_2,k}\left( t^{k}(k_1,k_2)\cdot\left(\left(\mathbf{M}^{\phantom{\dagger}}_{\Omega\to\Lambda K^-}S^{k_1}\left({\textstyle{\frac{3}{2}}},{\textstyle{\frac{3}{2}}}\right)\mathbf{M}^\dagger_{\Omega\to\Lambda K^-}\right)\otimes \left(\mathbf{M}^{\phantom{\dagger}}_{\bar{\Omega}\to\bar{\Lambda} K^+}S^{k_2}\left({\textstyle{\frac{3}{2}}},{\textstyle{\frac{3}{2}}}\right)\mathbf{M}^{\dagger}_{\bar{\Omega}\to\bar{\Lambda} K^+}\right)^{k}\right)\right),
\end{aligned}
\end{equation}
where the decay matrix $\mathbf{M}^{\phantom{\dagger}}_{\Omega\to\Lambda K^-}$ responsible for the decay of $\Omega$ has the invariant form
\begin{equation}
\label{M-Lambda}
\mathbf{M}^{\phantom{\dagger}}_{\Omega\to\Lambda K^-}=\sum_{\lambda=1,2}\left(Y_\lambda(\boldsymbol{\hat{p}_{{}_\Lambda}})\cdot S^\lambda\left({\textstyle{\frac{1}{2}}},{\textstyle{\frac{3}{2}}}\right)\right){\mathrm M}_\lambda,
\end{equation}
while that for the decay of the $\bar{\Omega}$ has the invariant form
\begin{equation}
\label{M-barLambda}
\mathbf{M}^{\phantom{\dagger}}_{\bar{\Omega}\to\bar{\Lambda}K^+}=\sum_{\lambda=1,2}\left(Y_\lambda(\boldsymbol{\hat{p}_{{}_{\bar{\Lambda}}}}))\cdot
S^\lambda\left({\textstyle{\frac{1}{2}}},{\textstyle{\frac{3}{2}}}\right)
\right){\mathrm M}_\lambda,
\end{equation}
where $\boldsymbol{\overrightarrow{p_{{}_\Lambda}}}$($\boldsymbol{\overrightarrow{p_{{}_{\bar\Lambda}}}}$) is the momentum of the $\Lambda$($\bar{\Lambda}$) in the rest frame of $\Omega$($\bar{\Omega}$) and ${\mathrm M}_\lambda$ (assumed to be the same for both $\Omega$ and $\bar{\Omega}$ decay) with $\lambda=1$ and $\lambda=2$ are respectively the parity-conserving and the parity-nonconserving decay amplitudes/constants.

Using eqs. (\ref{M-Lambda}) and (\ref{M-barLambda}) in eq. (\ref{rhoLambda}), we get, after standard Racah recoupling,
\begin{equation}
\rho_{\Lambda\bar{\Lambda}}=\sum_{k_1'',k_2''=0}^1\sum_{k''=|k_1''-k_2''|}^{(k_1''+k_2'')}\mathcal{
T}^{k''}(k_1'',k_2'')\cdot\left(S^{k_1''}\left({\textstyle{\frac{1}{2}}},{\textstyle{\frac{1}{2}}}\right)\otimes S^{k_2''}\left({\textstyle{\frac{1}{2}}},{\textstyle{\frac{1}{2}}}\right)\right)^{k''},
\end{equation}
where the Fano statistical tensors which completely characterize the final $\Lambda\bar{\Lambda}$ state are given by
\begin{equation}
\begin{aligned}
\label{Fano-Lambda}
&\mathcal{
T}^{k''}_{q''}(k_1'',k_2'')=\sum_{k_1,k_2=0}^3\sum_{k=|k_1-k_2|}^{(k_1+k_2)}\sum_{\lambda_1,\lambda_1',\lambda_2,\lambda_2'=1}^2\sum_{\lambda_1''=|\lambda_1-\lambda_1'|}^{(\lambda_1+\lambda_1')}\sum_{\lambda_2''=|\lambda_2-\lambda_2'|}^{(\lambda_2+\lambda_2')}
\sum_{k_1'=|\lambda_1-k_1|}^{(\lambda_1+k_1)}
\sum_{k_2'=|\lambda_2-k_2|}^{(\lambda_2+k_2)}
\sum_{\lambda'''=|\lambda_1''-\lambda_2''|}^{(\lambda_1''+\lambda_2'')}\\&
G(k_1,k_2,k,\lambda_1,\lambda_2,\lambda_1',\lambda_2',\lambda_1'',\lambda_2'',k_1',k_2',k_1'',k_2'',k'',\lambda'')\,\mathrm{M}_{\lambda_1}\mathrm{M}^*_{\lambda_1'}\mathrm{M}_{\lambda_2}\mathrm{M}^*_{\lambda_2'}\\&\times \left( t^k(k_1,k_2)\otimes \left( Y_{\lambda_1''}(\boldsymbol{\hat{p}_{{}_\Lambda}})\otimes Y_{\lambda_2''}(\boldsymbol{\hat{p}_{{}_{\bar{\Lambda}}}})\right)^{\lambda''}\right)^{k''}_{q''},
\end{aligned}
\end{equation}
where the geometrical factors
\begin{equation}
\begin{aligned}
&G(k_1,k_2,k,\lambda_1,\lambda_2,\lambda_1',\lambda_2',\lambda_1'',\lambda_2'',k_1',k_2',k_1'',k_2'',k'',\lambda'')=\left(\frac{2}{\pi}\right)(-1)^{k_1+k_1'-k_1''}[\lambda_1]^2[\lambda_1']^2[k_1']^2[k_1'']
\\&\times (-1)^{k_2+k_2'-k_2''}[\lambda_2]^2[\lambda_2']^2[k_2']^2[k_2''](-1)^{k+k''}[k_1][k_2][k][\lambda''] C(\lambda_1\lambda_1'\lambda_1'';000)C(\lambda_2\lambda_2'\lambda_2'';000)\\&\times W({\textstyle{\frac{3}{2}}}k_1{\textstyle{\frac{1}{2}}}\lambda_1;{\textstyle{\frac{3}{2}}}k_1')W({\textstyle{\frac{3}{2}}}k_2{\textstyle{\frac{1}{2}}}\lambda_2;{\textstyle{\frac{3}{2}}}k_2')W({\textstyle{\frac{1}{2}}}\lambda_1'{\textstyle{\frac{1}{2}}}k_1';{\textstyle{\frac{3}{2}}}k_1'')
W({\textstyle{\frac{1}{2}}}\lambda_2'{\textstyle{\frac{1}{2}}}k_2';{\textstyle{\frac{3}{2}}}k_2'')
\\&\times
W(\lambda_1k_1\lambda_1'k_1'';k_1'\lambda_1'')W(\lambda_2k_2\lambda_2'k_2'';k_2'\lambda_2'')
\left\{\begin{array}{lll}
\lambda_1'' & k_1'' & k_1 \\
\lambda_2'' & k_2'' & k_2 \\
\lambda'' & k'' & k
\end{array}\right\}.
\end{aligned}
\end{equation}
Note also that 
\begin{equation}
S^{1}_q\left({\textstyle{\frac{1}{2}}},{\textstyle{\frac{1}{2}}}\right)\equiv \sigma^1_q,
\end{equation}
where $\sigma^1_q$ are the irreducible components of the Pauli Spin operator, $\overrightarrow{\sigma}$.
\subsection{The second state of decay: the density matrix for $p\bar{p}$}
The density matrix $\rho_{p\bar{p}}$ is given by
\begin{equation}
\begin{aligned}
\label{rhop}
    \rho_{p\bar{p}}=\sum_{k_1'',k_2'',k''}\left(\mathcal{T}^{k''}(k_1'',k_2'')\cdot\left(\left(\mathbf{M}^{\phantom{\dagger}}_{\Lambda\to p \pi^-}S^{k_1''}\left({\textstyle{\frac{1}{2}}},{\textstyle{\frac{1}{2}}}\right)\mathbf{M}^\dagger_{\Lambda\to p \pi^-}\right)\otimes \left(\mathbf{M}^{\phantom{\dagger}}_{\bar{\Lambda}\to\bar{p} \pi^+}S^{k_2''}\left({\textstyle{\frac{1}{2}}},{\textstyle{\frac{1}{2}}}\right)\mathbf{M}^{\dagger}_{\bar{\Lambda}\to\bar{p} \pi^+}\right)^{k''}\right)\right),
\end{aligned}
\end{equation}
where the decay matrix $\mathbf{M}^{\phantom{\dagger}}_{\Lambda\to p \pi^-}$ responsible for the decay of $\Lambda$ has the invariant form
\begin{equation}
\label{M-p}
\mathbf{M}^{\phantom{\dagger}}_{\Lambda\to p \pi^-}=\sum_{\lambda=0,1}\left(Y_\lambda(\boldsymbol{\hat{p}_p})\cdot S^\lambda\left({\textstyle{\frac{1}{2}}},{\textstyle{\frac{1}{2}}}\right)\right){\mathcal M}_\lambda,
\end{equation}
and that for the decay of the $\bar{\Lambda}$ has the invariant form
\begin{equation}
\label{M-barp}
\mathbf{M}^{\phantom{\dagger}}_{\bar{\Lambda}\to\bar{p}\pi^+}=\sum_{\lambda=0,1}\left(Y_\lambda(\boldsymbol{\hat{p}_{\bar{p}}}))\cdot
S^\lambda\left({\textstyle{\frac{1}{2}}},{\textstyle{\frac{1}{2}}}\right)
\right){\mathcal M}_\lambda,
\end{equation}
where $\boldsymbol{\overrightarrow{p_p}}$($\boldsymbol{\overrightarrow{p_{\bar{p}}}}$) is the momentum of the $p$($\bar{p}$) in the rest frame of $\Lambda$($\bar{\Lambda}$) and ${\mathcal M}_\lambda$ with $\lambda=0$ and $\lambda=1$ are respectively the parity-nonconserving and the parity-conserving decay amplitudes.

Using eqs. (\ref{M-p}) and (\ref{M-barp}) in eq. (\ref{rhop}), we get
\begin{equation}
\rho_{p\bar{p}}=\sum_{k_3''',k_4'''=0}^1\sum_{k_5'''=|k_3'''-k_4'''|}^{(k_3'''+k_4''')}\mathrm{
T}^{k_5''}(k_3''',k_4''')\cdot\left(S^{k_3'''}\left({\textstyle{\frac{1}{2}}},{\textstyle{\frac{1}{2}}}\right)\otimes S^{k_4'''}\left({\textstyle{\frac{1}{2}}},{\textstyle{\frac{1}{2}}}\right)\right)^{k_5''},
\end{equation}
where the Fano statistical tensors which completely characterize the final state are given by
\begin{equation}
\begin{aligned}
\label{Fano-p}
&\mathrm{
T}^{k_5'''}_{q_5'''}(k_3''',k_4''')=\sum_{k_1'',k_2'',k_3'',k_4''=0}^1\sum_{k''=|k_1''-k_2''|}^{(k_1''+k_2'')}\sum_{\lambda_3,\lambda_3',\lambda_4,\lambda_4'=0}^1\sum_{\lambda_3''=|\lambda_3-\lambda_3'|}^{(\lambda_3+\lambda_3')}\sum_{\lambda_4''=|\lambda_4-\lambda_4'|}^{(\lambda_4+\lambda_4')}
\sum_{\lambda_5''=|\lambda_3''-\lambda4''|}^{(\lambda_3''+\lambda_4'')}
\\&
\mathcal{G}(k_1'',k_2'',k'',\lambda_3,\lambda_4,\lambda_3',\lambda_4',\lambda_3'',\lambda_4'',\lambda_5'',k_3'',k_4'',k_3''',k_4''',k_5''')\,\mathcal{M}_{\lambda_3}\mathcal{M}^*_{\lambda_3'}\mathcal{M}_{\lambda_4}\mathcal{M}^*_{\lambda_4'}\\&\times \left(\mathcal{T}^{k''}(k_1'',k_2'')\otimes \left( Y_{\lambda_3''}(\boldsymbol{\hat{p}_{p}})\otimes Y_{\lambda_4''}(\boldsymbol{\hat{p}_{\bar{p}}})\right)^{\lambda_5''}\right)^{k_5'''}_{q_5'''},
\end{aligned}
\end{equation}
where the geometrical factors
\begin{equation}
\begin{aligned}
&\mathcal{G}(k_1'',k_2'',k'',\lambda_3,\lambda_4,\lambda_3',\lambda_4',\lambda_3'',\lambda_4'',\lambda_5'',k_3'',k_4'',k_3''',k_4''',k_5''')=\left(\frac{1}{\pi}\right)(-1)^{\lambda_3+k_3''-k_3'''}[\lambda_3]^2[\lambda_3']^2[k_3'']^3
\\&\times (-1)^{\lambda_4+k_4''-k_4'''-k_5'''}[\lambda_4]^2[\lambda_4']^2[k_4'']^3[\lambda_5''][k_1''][k_2'']
 C(\lambda_3\lambda_3'\lambda_3'';000)C(\lambda_4\lambda_4'\lambda_4'';000)\\&\times W({\textstyle{\frac{1}{2}}}k_1''{\textstyle{\frac{1}{2}}}\lambda_3;{\textstyle{\frac{1}{2}}}k_3'')W({\textstyle{\frac{1}{2}}}k_2''{\textstyle{\frac{1}{2}}}\lambda_4;{\textstyle{\frac{1}{2}}}k_4'')W({\textstyle{\frac{1}{2}}}\lambda_3'{\textstyle{\frac{1}{2}}}k_3'';{\textstyle{\frac{1}{2}}}k_3''')
W({\textstyle{\frac{1}{2}}}\lambda_4'{\textstyle{\frac{1}{2}}}k_4'';{\textstyle{\frac{1}{2}}}k_4''')
\\&\times
W(\lambda_3k_1''\lambda_3'k_3''';k_3''\lambda_3'')W(\lambda_4k_2''\lambda_4'k_4''';k_4''\lambda_4'')
\left\{\begin{array}{lll}
\lambda_3'' & k_3''' & k_1'' \\
\lambda_4'' & k_4''' & k_2'' \\
\lambda_5'' & k_5''' & k''
\end{array}\right\}.
\end{aligned}
\end{equation}
Using eq. (\ref{Fano-Lambda} in eq. (\ref{Fano-p}), we can relate the Fano statistical tensors of the $\Omega\bar{\Omega}$ to that of $p\bar{p}$ as
\begin{equation}
\begin{aligned}
\label{Fano-final}
\mathrm{
T}^{k_5'''}_{q_5'''}(k_3''',k_4''')&=\sum \mathcal{G}(k_1'',k_2'',k'',\lambda_3,\lambda_4,\lambda_3',\lambda_4',\lambda_3'',\lambda_4'',\lambda_5'',k_3'',k_4'',k_3''',k_4''',k_5''')\\&\times G(k_1,k_2,k,\lambda_1,\lambda_2,\lambda_1',\lambda_2',\lambda_1'',\lambda_2'',k_1',k_2',k_1'',k_2'',k'',\lambda'')\\&\times \mathrm{M}_{\lambda_1}\mathrm{M}^*_{\lambda_1'}\mathrm{M}_{\lambda_2}\mathrm{M}^*_{\lambda_2'}\mathcal{M}_{\lambda_3}\mathcal{M}^*_{\lambda_3'}\mathcal{M}_{\lambda_4}\mathcal{M}^*_{\lambda_4'} \Bigl(\left( t^k(k_1,k_2)\otimes \left( Y_{\lambda_1''}(\boldsymbol{\hat{p}_{{}_\Lambda}})\otimes Y_{\lambda_2''}(\boldsymbol{\hat{p}_{{}_{\bar{\Lambda}}}})\right)^{\lambda''}\right)^{k''}\\&\otimes \left( Y_{\lambda_3''}(\boldsymbol{\hat{p}_{p}})\otimes Y_{\lambda_4''}(\boldsymbol{\hat{p}_{\bar{p}}})\right)^{\lambda_5''}\Bigr)^{k_5'''}_{q_5'''}.
\end{aligned}
\end{equation}
The orthogonality of the $Y_{lm}$s allows us to ``invert'' the above equation, allowing us to write the above equation, as
\begin{equation}
\begin{aligned}
\label{int-Fano}
&t^{k_5'''}_{q_5'''}(k_3''',k_4''')=\Bigl[(16\pi^2)\sum \mathcal{G}(k_3''',k_4''',k_5''',\lambda_3,\lambda_4,\lambda_3,\lambda_4,0,0,0,k_3'',k_4'',k_3''',k_4''',k_5''')\\&\times G(k_3''',k_4''',k_5''',\lambda_1,\lambda_2,\lambda_1,\lambda_2,0,0,k_1',k_2',k_3''',k_4''',k_5''',0)|\mathrm{M}_{\lambda_1}|^2|\mathrm{M}_{\lambda_2}|^2|\mathcal{M}_{\lambda_3}|^2|\mathcal{M}_{\lambda_4}|^2\Bigr]^{-1}\\&\times
\int\mathrm{
T}^{k_5'''}_{q_5'''}(k_3''',k_4''')d\Omega_{\boldsymbol{\hat{p}_{{}_\Lambda}}}d\Omega_{\boldsymbol{\hat{p}_{{}_{\bar{\Lambda}}}}}d\Omega_{\boldsymbol{\hat{p}_{p}}}d\Omega_{\boldsymbol{\hat{p}_{\bar{p}}}}.
\end{aligned}
\end{equation}
\section{Results and discussion}
If the decay amplitudes are known, the above equation will allow us to ``solve'' for some of the Fano-statistical tensors of the original $\Omega\bar{\Omega}$ system, up to $k_5'''=2$, using the Fano-statistical tensors for the final $p\bar{p}$ system.  Most importantly, the hitherto ``unknown'' production (differential) cross section \cite{PANDA} for $\Omega\bar{\Omega}$ production can be determined by putting $k_3'''=k_4'''=k_5'''=q_5'''=0$ in eq. (\ref{int-Fano}, which implies that the geometrical factors now reduce to
\begin{equation}
\mathcal{G}(0,0,0,\lambda_3,\lambda_4,\lambda_3,\lambda_4,0,0,0,\lambda_3,\lambda_4,0,0,0)=\frac{1}{4\pi}[\lambda_3]^3[\lambda_4]^3
\end{equation}
and
\begin{equation}
G(0,0,0,\lambda_1,\lambda_2,\lambda_1,\lambda_2,0,0,\lambda_1,\lambda_2,0,0,0,0)=\frac{1}{4\pi}(-1)^{\lambda_1+\lambda_2}[\lambda_1]^2[\lambda_2]^2,
\end{equation}
so that the production (differential) cross section for $\bar{p}p\to \bar{\Omega}\Omega$, which is essentially eq. (\ref{unpol-dcs2}), is given by
\begin{equation}
\begin{aligned}
t^0_0(0,0)&=\Bigl[\sum_{\lambda_1,\lambda_2,\lambda_3,\lambda_4}(-1)^{\lambda_1+\lambda_2}[\lambda_1]^2[\lambda_2]^2[\lambda_3]^3[\lambda_4]^3|\mathrm{M}_{\lambda_1}|^2|\mathrm{M}_{\lambda_2}|^2|\mathcal{M}_{\lambda_3}|^2|\mathcal{M}_{\lambda_4}|^2\Bigr]^{-1}\\&\times\int\mathrm{
T}^{0}_{0}(0,0)d\Omega_{\boldsymbol{\hat{p}_{{}_\Lambda}}}d\Omega_{\boldsymbol{\hat{p}_{{}_{\bar{\Lambda}}}}}d\Omega_{\boldsymbol{\hat{p}_{p}}}d\Omega_{\boldsymbol{\hat{p}_{\bar{p}}}}.
\end{aligned}
\end{equation}
The above result demonstrates that the model-independent irreducible tensor formalism we have developed here would be of immense use for upcoming facilities like PANDA, where kinematically complete experiments are being planned to analyse sequential decays of these triple strange process.

\begin{acknowledgments}
The author acknowledges with gratitude the financial support received under the DST-SERB grant, CRG/2021/000435.
\end{acknowledgments}

\end{document}